# Interference and switching effect of topological interfacial modes with geometric phase

Xing-Xiang Wang[1,2], Tomohiro Amemiya[3], and Xiao Hu[1,2]*

[1]Research Center for Materials Nanoarchitectonics (MANA), National Institute for Materials Science (NIMS); Tsukuba 305-0044, Japan.

[2]Graduate School of Science and Technology, University of Tsukuba; Tsukuba 305-8571, Japan.

[3]School of Engineering, Tokyo Institute of Technology; Tokyo 152-8552, Japan.

*Corresponding author's Email: Hu.Xiao@nims.go.jp

## Abstract

We investigate interference between topological interfacial modes in a semiconductor photonic crystal platform with Dirac frequency dispersions, which can be exploited for interferometry switch. It is showcased that, in a two-in/two-out structure with four topological waveguides, geometric phases of the two-component spinor wavefunctions of topological photonic modes accumulate at turning points of waveguides, which govern the interferences and split the electromagnetic energy into two output ports with relative power ratio tunable by the relative phase of inputs. We unveil that this brand-new photonic phenomenon is intimately related to the spin-momentum locking property of quantum spin Hall effect, and results from the symphonic contributions of three phase variables: the spinor phase and geometric phase upon design, and the global phase controlled from outside. The present findings open the door for manipulating topological interfacial modes, thus exposing

**a new facet of topological physics. The topology-driven interference can be incorporated into other devices which is expected to leave far-reaching impacts to advanced photonics, optomechanics and phononics applications.**

## Introduction

The information society flourishing nowadays is supported largely by fast data processing based on electronics [1, 2] and high-velocity data transfer relying on optics [3]. As is well known, both transport of electrons in integrated circuits (IC) [2] and transformation of signal from electrons to light and vice versa induce heating, which becomes more serious when devices are downsizing and processing speed increases. Light moves faster than electrons and meanwhile does not cause heat, and thus processing information in terms of light is more efficient in principle. Therefore, to reduce the times of transformation between optic and electric signals, and to increase the usage of light in the stream of information processing are highly desirable. In the growing fields such as artificial intelligence (AI) technology [4, 5], the advantage of light data processing is even more obvious [6, 7]. The demand on replacing electron by light for information processing would be tremendous in the coming years. However, to harness light is far more difficult than to control electrons, and realizing efficient optic splitter, switch and isolator is highly yearned for advanced data processing based on optic IC.

Photonics topology provides a completely new possibility for controlling light propagation [8, 9, 10, 11, 12, 13, 14, 15, 16, 17, 18, 19, 20, 21, 22, 23]. By now topological waveguide propagation robust against disorder, randomness and sharp corners have been displayed, which breaks the limitation of conventional electromagnetic (EM) wave. As an additional ingredient of topology of bosonic systems, topological interfacial modes have been shown to sustain stable lasing superior to conventional whispering-gallery-modes [24, 25, 26]. Moreover, wavefunction partition of one topological interfacial mode into two topological channels has also been investigated which depends on the mutual angles and/or pseudospin states [27, 28, 29, 30, 31, 32, 33]. On the other hand, interference of topological interfacial/edge modes piloted by phases remains largely unexplored so far for both fermionic and bosonic systems.

Here, we explore interference between topological interfacial modes in a semiconductor photonic crystal (PhC) platform. The photonics topology emerges from Kekulé distortions respecting $C_{6v}$ symmetry to honeycomb structure with Dirac frequency dispersions, where gapless helical interfacial modes protected by the mirror and sublattice symmetries appear along the "molecule zigzag" interface between topological PhC and trivial PhC [17, 34]. We demonstrate that, in a two-in/two-out structure with four topological waveguides, geometric phases of the two-component spinor wavefunctions of topological photonic modes accumulate along the designed waveguides, which split the EM energy into two output ports with relative power ratio tunable by the relative phase of inputs. This photonic phenomenon is intimately related to the spin-momentum locking property of quantum spin Hall effect (QSHE), and results from the symphonic contributions of three optic phases: the two-component spinor phase, geometric phase and global phase, where the first two are upon design and the last one is the handle controllable from outside. This brand-new photonic property is realized due to a unique interplay between the intrinsic bosonic feature of light and the emergent fermionic property encoded in the spin-1/2 spinor wavefunction in the semiconductor topological waveguide system, which on one hand enables an interferometry switch and on the other hand exposes a new facet of topology physics.

**Results**

The frequency dispersion of the honeycomb-type PhC with $C_{6v}$ symmetric Kekulé distortions displayed in Fig. 1a is described by the Dirac-type $\mathbf{k} \cdot \mathbf{p}$ Hamiltonian on the basis $\{|p_+\rangle, |d_+\rangle, |p_-\rangle, |d_-\rangle\}$ [13, 35]

$$\hat{H} = \begin{bmatrix} -M & -ivk_+ & 0 & 0 \\ ivk_- & M & 0 & 0 \\ 0 & 0 & -M & -ivk_- \\ 0 & 0 & ivk_+ & M \end{bmatrix}, \quad (1)$$

with $M$ the Dirac mass, $v > 0$ the Dirac velocity, $k_\pm = k_x \pm ik_y$ where $\mathbf{k} = (k_x, k_y)$ is measured from Γ point of the Brillouin zone of the hexagonal unit cell, and the basis wavefunctions referring to the field $H_z$ in the hexagonal unit cell for the TE mode. Since the topological interface states with frequency around the bandgap center are under concern in the present work, high-order terms $O(k^2)$ in the $\mathbf{k} \cdot \mathbf{p}$ theory can be neglected as a sufficient approximation, which renders Hamiltonian (1) block diagonalized into the pseudospin-up and -down subspaces (see also [26, 36]). It was shown that the Dirac mass in Eq. (1) takes a positive (negative) value for the trivial (topological) PhC structure with $a_0/R < 3$ ($a_0/R > 3$) as shown in Fig. 1b (Fig. 1c), characterized by the $p - d$ band inversion [13].

In order to unveil the functioning principle of the topological interferometry switch, we begin with the topological waveguide mode along a horizontal interface where the topological/trivial PhC occupies the upper/lower half space respectively, as shown for the input Port A in Fig. 1a and schematically displayed in Fig. 2a. For the pseudospin-up sector, one has from Eq. (1)

$$\hat{H}^+ = \begin{bmatrix} -M(y) & -iv\partial_y \\ -iv\partial_y & M(y) \end{bmatrix} \quad (2)$$

where $k_y = -i\partial_y$ is plugged in, which gives the following solution associated with a zero eigenvalue, i.e. frequency at the center of bandgap

$$\psi(y) = \begin{bmatrix} 1 \\ i \end{bmatrix} e^{yM(y)/v} \tag{3}$$

where $M(y) = m > 0$ for $y < 0$ and $M(y) = -m < 0$ for $y > 0$ (the Dirac mass is presumed to take the same absolute value $m$ in the two PhCs for simplicity). This is the photonic realization of the Jackiw-Rebbi zero-energy soliton solution originally proposed for fermion number 1/2 field [37]. With the basis taken into account, the full real-space solution of the zero-energy pseudospin-up state is

$$|0, +\rangle = [|p_+\rangle \, |d_+\rangle] \begin{bmatrix} 1 \\ i \end{bmatrix} e^{yM(y)/v}. \tag{4}$$

Due to the interference between the $|p_+\rangle$-mode and $|d_+\rangle$-mode carrying counterclockwise phase winding with vorticity +1 and +2 respectively (see Fig. 2b and c), there appears a net EM energy flow along the $+x$ direction as displayed in Fig. 2d, which is consistent with the group velocity of pseudospin-up states [13].

As the time-reversal symmetry counterpart, there is a state supported by the pseudospin-down basis which transports a net EM energy flow along the $-x$ direction. In the setup shown in Fig. 2a where the EM signal is injected from the left to the system (corresponding to the input Port A in Fig. 1a), only the pseudospin-up subspace is relevant due to the spin-momentum locking in the present photonic analogue of QSHE. Keep in mind that in topological insulators spin-down electrons go around the outer sample edge and cannot be excluded from the whole interference process, in a sharp contrast to the present photonic system [44, 45, 49, 50].

Next, let us consider the EM wavefunction of the topological interfacial waveguide in the form of an arc between two straight segments, one along the $+x$ direction as discussed above and the other takes an angle $\theta$ measured from the $x$ axis (see Fig. 2a). In this case, a polar-like coordinate $(\theta, \rho)$ is convenient at the arc segment with radius $\mathcal{R}$ as displayed in Fig. 2a (straight interfaces can be considered formally as the limit $\mathcal{R} \to \infty$). Here one has

$$k_\pm = k_x \pm i k_y = e^{\pm i\theta}(k_\theta \pm i k_\rho), \tag{5}$$

with the momentum operators

$$(k_\theta, k_\rho) = (-i\partial_\theta/\mathcal{R}, -i\partial_\rho). \tag{6}$$

We separate the pseudospin-up sector of Hamiltonian (1) into two parts

$$\widehat{H}^+ = \begin{bmatrix} -M & -ivk_+ \\ ivk_- & M \end{bmatrix} = \widehat{H}_0^+ + \widehat{H}_1^+ \tag{7}$$

with

$$\widehat{H}_0^+ = \begin{bmatrix} -M & ve^{i\theta}k_\rho \\ ve^{-i\theta}k_\rho & M \end{bmatrix} = \begin{bmatrix} -M & -ive^{i\theta}\partial_\rho \\ -ive^{-i\theta}\partial_\rho & M \end{bmatrix} \tag{8}$$

and

$$\widehat{H}_1^+ = \begin{bmatrix} 0 & -ive^{i\theta}k_\theta \\ ive^{-i\theta}k_\theta & 0 \end{bmatrix} = \frac{1}{\mathcal{R}}\begin{bmatrix} 0 & -ve^{i\theta}\partial_\theta \\ ve^{-i\theta}\partial_\theta & 0 \end{bmatrix}, \tag{9}$$

where $\widehat{H}_0^+$ generates the zero-energy soliton state same as Eq. (2), while $\widehat{H}_1^+$ is responsible for the bending of the interface waveguide which is to be treated as a perturbation [52]. Hamiltonian (8) is solved in the way same as Eq. (2), which yields the zero-energy wavefunction

$$|\theta, +\rangle = [|p_+\rangle |d_+\rangle]\begin{bmatrix} 1 \\ ie^{-i\theta} \end{bmatrix}F(\rho), \tag{10}$$

with $F(\rho)$ a function taking maximum at the interface ($\rho = 0$) and decaying exponentially into the bulks in the way shown in Eq. (4). It is obvious that Eq. (10) gives the wavefunction of the topological interface mode along the straight interface in direction of $\theta$ (see Fig. 2a). As can be seen clearly in Fig. 2e, f and g for $\theta = 2\pi/3, \pi$ and $4\pi/3$, respectively, which are compatible with the PhC structure up to the length scale of lattice constant $a_0$, the phase factor in the two-component spinor governs the interference between $|p_+\rangle$-mode and $|d_+\rangle$-mode within the unit cell, which yields the net EM energy flow along the interface between the topological and trivial PhCs in the same way as Eq. (4).

Now we treat $\widehat{H}_1^+$ in Eq. (9) by the first-order perturbation where $|\theta, +\rangle$ in Eq. (10) serves as the unperturbed wavefunction

$$\widetilde{H}_1^+ = \langle \theta, +|\widehat{H}_1^+|\theta, +\rangle = \hbar\omega_0 \left(-i\partial_\theta - \frac{1}{2}\right), \tag{11}$$

with $\hbar\omega_0 = (2v/\mathcal{R}) \int d\rho F(\rho)^2$, where the factor $1/2$ in the last result of Eq. (11) appears due to the commutation relation and the Dirac Hamiltonian. One then has the eigenvalue equation for $\widetilde{H}_1^+$

$$\hbar\omega_0 \left(-i\partial_\theta - \frac{1}{2}\right)\psi(\theta) = \hbar\omega\psi(\theta), \tag{12}$$

where $\psi(\theta)$ is the coefficient of the unperturbed wavefunction $|\theta, +\rangle$. For the topological interface waveguide mode at $\omega = 0$ (i.e. the frequency at bandgap center), the solution for Eq. (12) is

$$\psi(\theta) = \psi(\theta_{\text{in}}) \exp[i(\theta - \theta_{\text{in}})/2], \tag{13}$$

with $\theta_{\text{in}}$ the direction of topological interface waveguide where the arc segment starts.

Equation (13) indicates that as the Jackiw-Rebbi mode $|\theta, +\rangle$ propagates along an arc segment in the topological interface waveguide, it acquires a phase accumulation corresponding to half of angle of arc. Noticing that at $\omega = 0$, the arc radius $\mathcal{R}$ becomes irrelevant in Eq. (12), which implies that Eq. (13) is applicable to a sharp turn with $\mathcal{R} \sim a_0$ (as will be confirmed by numerical simulations below). It is straightforward to see that Eq. (13) is also applicable to the topological interface waveguide bending to the side of trivial PhC.

Now we consider the design displayed in Fig. 1a, where four domains of topological and trivial PhCs form four topological interface waveguides with a crossing point. Owing to the pseudospin-momentum locking associated with the $Z_2$ topology, the beam from input Port A cannot get into the interface channel between the crossing point and the input Port B since along this direction of

EM flow a pseudospin-down state is required where the topological/trivial PhCs are turned over (namely the topological/trivial PhCs occupy the lower/upper half space). Therefore, the beam injected from Port A is split into the two output channels, Port 1 and Port 2, with the EM energy equally divided as guaranteed by the symmetry of the topological waveguide system in Fig. 1a.

Similarly, the EM wave injected from the input Port B propagates along the topological interface waveguide with a pseudospin-up state, which cannot go beyond the crossing point into the interface channel toward input Port A, and is split equally in energy into the two output Port 1 and Port 2. Therefore, as a crucial feature of the present topological interface waveguide system, the two input beams from Port A and Port B do not form a standing wave in the horizontal interfaces as guaranteed by the pseudospin-momentum locking originated from the $Z_2$ photonics topology, which is not available in conventional optical waveguide systems.

It is easy to see that the wavefunction along the straight interface from input Port B is also given by Eq. (10) with $\theta = \pi$, and Eq. (13) is also available when the topological interface waveguide is bent. From Eq. (10) it is clear that the wavefunctions $|0,+\rangle$ and $|\pi,+\rangle$ are orthogonal to each other (see also Fig. 2d and f), which corresponds to the pseudospin-momentum locking effect in the above discussions.

In order to verify the above properties, we have checked the branching of the EM beams injected from Port A at the crossing point in terms of numerical simulations based on COMSOL [51] (see Methods). It is seen clearly in Fig. 3a, where the wavefunction defined by the out-of-plane magnetic field $H_z$ is displayed for the present TE mode, that the EM beam is split into Port 1 and Port 2 with approximately equal weights irrespectively to the global phase determined by the chiral source, and it is prevented from propagating beyond the crossing point into the channel toward input Port B. Moreover, as displayed in Fig. 3c, d and e, one can find that phases along the straight interface segments remain almost constant, since the frequency of chiral source is chosen at the bandgap center which corresponds to a zero Bloch phase upon propagation, in good agreement with theoretical description given in Eq. (10) [see also Eq. (4)]. Noticeably, phase changes are accumulated upon bending into output Port 1 and Port 2, which take place within a couple of unit cells around the crossing point. These simulation results confirm the theoretical result of Eq. (13). Similarly, one can find the same physics happens when the input Port B is excited, as shown in Fig. 3b and i~n.

We then proceed to analyze the interference of topological interface modes along the two output channels in the design displayed in Fig. 1a when EM waves are injected simultaneously from the input Port A and Port B. It is clear that the output wavefunctions are the superpositions of the wavefunctions contributed by input Port A and Port B individually. Suppose that the wavefunctions injected from the two input ports are $e^{i\phi}|0,+\rangle$ at Port A and $|\pi,+\rangle$ at Port B, where $\phi$ is the global phase difference between the two input ports which can be controlled from outside, and the output Port 1 (Port 2) is along the direction of $\theta_1 = 2\pi/3$ ($\theta_2 = 4\pi/3$) as given in Fig. 1a. According to Eqs. (10) and (13), the output wavefunctions at the two output ports are given by

$$|1\rangle_{\text{total}} = |1\rangle_A + |1\rangle_B = e^{i(\theta_1-\pi)/2}\left[1 + e^{i\left(\phi+\frac{\pi}{2}\right)}\right]|\theta_1,+\rangle, \tag{14}$$

and

$$|2\rangle_{\text{total}} = |2\rangle_A + |2\rangle_B = e^{i(\theta_2-\pi)/2}\left[1 + e^{i\left(\phi-\frac{\pi}{2}\right)}\right]|\theta_2,+\rangle, \tag{15}$$

where the strengths of the two input EM waves are put same. The output powers at output Port 1 and Port 2 normalized by the total input power are then evaluated as

$$P_1 = \frac{\langle 1|1\rangle_{\text{total}}}{\langle 1|1\rangle_{\text{total}} + \langle 2|2\rangle_{\text{total}}} = \frac{1}{2}(1 - \sin\phi),$$

$$P_2 = \frac{\langle 2|2\rangle_{\text{total}}}{\langle 1|1\rangle_{\text{total}} + \langle 2|2\rangle_{\text{total}}} = \frac{1}{2}(1 + \sin\phi). \tag{16}$$

Equation (16) shows explicitly that, by adjusting the phase difference $\phi$ between the two input ports, one can manipulate continuously powers at the two output ports, as shown in Fig. 4a. Especially at $\phi = -\pi/2$, the output is directed totally to Port 1 and turned off at Port 2, and vice versa at $\phi = \pi/2$. This is the functioning principle of the interferometry switch based on the topological interfacial waveguides.

The switching effect can also be confirmed by numerical simulations, where the input Port A and Port B are simultaneously stimulated and the relative phase between the two inputs is tuned. Figure 4a clearly illustrates that powers at output Port 1 and Port 2 exhibit a sinusoidal variation upon tuning $\phi$, which is in excellent agreement with Eq. (16) derived from theoretical analysis. We display in Fig. 4b-f the distribution of intensity of the EM wave in terms of $|H_z|^2$ at several values of $\phi$ to visualize the switching process. At $\phi = -\pi/2$, the output is almost entirely directed to Port 1. As $\phi$ increases to $\phi = -\pi/4$, there is a slight transfer of power to output Port 2. At $\phi = 0$, the two output ports exhibit nearly equal powers. When $\phi = \pi/4$, the output is primarily directed to Port 2. Finally, at $\phi = \pi/2$, the output power at Port 1 is suppressed almost completely. The whole switching process occurs due to the interference of the wavefunctions contributed by Port A and Port B, as indicated in the theoretical discussions.

It is worthy noticing that the nontrivial phase accumulation encoded in the factor of 1/2 in Eq. (13) upon bending of the topological interface waveguide along an arc is crucial for the switching effect given in Eq. (16). As seen in Eqs. (14) and (15), the relative direction change of beams injected from input Port A and Port B is of $\pi$ for the EM wavefunction in output Port 1, and of $-\pi$ for that output Port 2. The reason why they can generate EM energy flows different in the two output ports is that the phase accumulations induced by the direction changes yield phase factors $e^{i\pi/2}$ and $e^{-i\pi/2}$ unequal to each other in the last equality in Eqs. (14) and (15) (without the factor 1/2 one would have $e^{i\pi} = e^{-i\pi}$).

**Discussions**

**Geometric phase**

The switching effect addressed above has a clear underlying physics of geometric phase [53, 54] associated with the Dirac Hamiltonian. Considering the basis wavefunction $|\theta, +\rangle = [1 \quad ie^{-i\theta}]^T/\sqrt{2}$ in Eq. (10), which adiabatically evolves in the one-dimensional (1D) parametric space $\theta$, the geometric phase accumulated in the waveguide bent from $\theta_1$ to $\theta_2$ is evaluated as [53, 54]

$$\gamma(\theta_1, \theta_2) = \int_{\theta_1}^{\theta_2} d\theta \langle \theta, +|i\partial_\theta|\theta, +\rangle = \frac{1}{2}\int_{\theta_1}^{\theta_2} d\theta = (\theta_2 - \theta_1)/2, \quad (17)$$

which gives directly Eq. (13). While the above result is gauge dependent since the path is not closed, it is clear that the relation Eq. (13) for $\psi(\theta)$ is gauge independent since it refers to the measurable physical quantity. During the derivation of Eq. (13), we perceive the interface bending as a perturbation, and require the wavefunction to stay at zero energy under this perturbation. This treatment corresponds to the above adiabatic evolution without altering the energy level of the eigenstate.

**Spin-1/2 degree of freedom**

The phenomena discussed above can also be captured in terms of spin-1/2 degree of freedom of the topological interface mode. The block diagonalized Dirac Hamiltonian (1) governs the two-component spinor on basis $\{|p_+\rangle, |d_+\rangle\}$ characterized by a fermionic field with $s = 1/2$ [37]. From Eq. (10), one has $\langle \sigma_x \rangle = \sin\theta$ and $\langle \sigma_y \rangle = \cos\theta$, and the EM beam injected from input Port

A and Port B corresponds to state $\langle \sigma_y \rangle = 1$ and $\langle \sigma_y \rangle = -1$, respectively ($\langle \sigma_x \rangle = 0$ in both cases). Upon conflowing into output Port 1 (Port 2) the spin rotates by $\delta\theta = \pi \, (-\pi)$ and induces relative phase factor $e^{is\delta\theta} = e^{i\pi/2} \, (e^{-i\pi/2})$, which contribute to Eqs. (14) and (15) with opposite signs. On the other hand, the injection of EM beams from the input Port A and Port B toward the crossing point of topological interfacial waveguides is certainly inherited from the Bose-Einstein statistics of photons, which induces interferences along the output channels. Therefore, the topological switching effect uncovered in the present work is due to the unique interplay between the intrinsic bosonic feature of light and the emergent fermionic property encoded in the spin-1/2 spinor wavefunction in the semiconductor topological waveguide system.

**Zero-energy mode**

In the present work, we adopt the design based on the so-called "molecule zigzag" interface, where the existence of the zero-energy topological interfacial mode is guaranteed by the mirror symmetry and chiral (i.e. sublattice) symmetry [34] (see also experiments in [17]). We have also confirmed that all the results presented here remain valid in a design based on the "armchair" interface, where the tiny minigap in dispersions of topological interfacial modes due to the weak crystalline symmetry breaking at interfaces [13] is naturally smeared out by the finite system size in simulations as well as in practical devices.

**Beam splitting of one input vs interference between two inputs**

Wavefunction partition, or beam splitting, of topological interfacial modes has been addressed in previous works [27, 28, 29, 30, 31, 32, 33], which is unveiled to depend on the mutual angles between one input port and two output ports. In stark contrast, in the present work we design the topological waveguide system in such a way that the two output ports are symmetric respect to the input ports, which guarantees the equal energy partition when only one input port is fed. The topological switch effect toward two output ports is purely due to the interference between the two topological input beams, which is piloted by the relative phase difference of the two input ports.

**Device applications**

The key interferences of topological interfacial modes take place around the crossing point within several unit cells, with the length scale of ~3 μm at the telecommunication frequency, which sets the limiting device size of the present topological switch (see simulation results in Fig. 4a and b). This structure may also be incorporated into other Mach-Zehnder type devices [55].

The present topological switch can be easily implanted into silicon nanophotonic systems, since it has been demonstrated experimentally that the conventional silicon waveguides with simple TE/TM modes can be smoothly converted into the topological interfacial waveguides and vice versa [56], which therefore can work as the low-loss input and output ports for the present topological interferometry switch. The optical switching effect addressed in the present study can work as building block for other key elements of light IC, such as diode and transistors [1].

The interference and switching functions of the topological interfacial waveguide modes unveiled in the present work are available for other wave systems with intrinsic bosonic feature, including optomechanical, surface acoustic and phononic waves [22, 57, 58]. Applications of the basic idea of the present approach to electronic systems, such as exciton-polariton structures, are also anticipated [20]. All these are expected to boost discontinuous leaps in exploiting matter topology for brand-new phenomena and innovative technologies.

**Methods**

All the numerical simulations presented in this study were performed using the Finite Element Method (FEM) implemented in COMSOL Multiphysics software [51]. The topological PhC platform in the present work is considered as a dielectric slab with hexagonal unit cells each containing six airhole and respecting $C_{6v}$ symmetry (see Fig. 1 in the main text). In order to reduce computation resource, we emulate the topological waveguide switch device by considering physical parameters and EM fields uniform in the $z$ direction, which transfers numerical calculations into purely 2D. For a silicon slab with thickness ~200 nm, the relative permittivity is taken as $\epsilon_r = 7$. The lattice constant of the PhC is $a_0 = 800$ nm, with each hexagonal unit cell containing six triangular airholes, and the side length of each airhole is $L = 270$ nm. In the case of the trivial PhC, the distance from the center of unit cell to the center of airholes is $R = 245$ nm ($< a_0/3$), while for the topological PhC, this distance is $R = 280$ nm ($> a_0/3$). The total topological switch in the simulations (see Fig. 3 and Fig. 4) consists of 47 unit cells along the longer side and 24 unit cells along the shorter side, and is separated into four regions by the interfaces between the topological and trivial domains. A chiral source is placed at the center of a unit cell at the end of each input port, formed by a cluster of six point-like sources arranged in the shape of a small hexagon with a $2\pi$ phase winding (a phase difference of $\pi/3$ between neighboring point sources) in the counterclockwise direction, which induce the pseudospin-up topological interfacial modes in the system.

## Data availability

All the data that support the findings of this study are available from the corresponding author upon reasonable request, following the policy of JST.

## Code availability

All the codes that support the findings of this study are available from the corresponding authors upon reasonable request, following the policy of JST.

# References

1. Huff, H. R. John Bardeen and transistor physics. *AIP Conference Proceedings* **550**, 3 (2001).

2. Horowitz, P. & Hill, W. *The art of electronics*. (Cambridge University Press, 1989).

3. Kao, C. K. *Optical fiber technology, II*. (Institute of Electrical and Electronics Engineers: sole worldwide distributor, Wiley, 1981).

4. Russell, S. J. & Norvig, P. *Artificial intelligence: a modern approach*. (Pearson, 2021).

5. Rich, E. & Knight, K. *Artificial intelligence (3rd Edition)*. (McGraw-Hill-India, 2010).

6. Shen, Y., Harris, N. C., Skirlo, S., Prabhu, M., Baehr-Jones, T., Hochberg, M., Sun, X., Zhao, S., Larochelle, H., Englund, D. & Soljačić, M. Deep learning with coherent nanophotonic circuits. *Nat. Photon.* **11**, 441–446 (2017).

7. Chen, Z., Sludds, A., Davis, R., Christen, I., Bernstein, L., Ateshian, L., Heuser, T., Heermeier, N., Lott, J. A., Reitzenstein, S., Hamerly, R. & Englund, D. Deep learning with coherent VCSEL neural networks. *Nat. Photon.* **17**, 723 (2023).

8. Haldane, F. D. M. & Raghu, S. Possible Realization of Directional Optical Waveguides in Photonic Crystals with Broken Time-Reversal Symmetry. *Phys. Rev. Lett*. **100**, 013904 (2008).

9. Wang, Z., Chong, Y., Joannopoulos, J. D. & Soljacic, M. Observation of unidirectional backscattering-immune topological electromagnetic states. *Nature* **461**, 772 (2009).

10. Hafezi, M., Mittal, S., Fan, J., Migdall, A. & Taylor, J. M. Imaging topological edge states in silicon photonics. *Nat. Photon*. **7**, 1001 (2013).

11. Rechtsman, M. C., Zeuner, J. M., Plotnik, Y., Lumer, Y., Podolsky, D., Dreisow, F., Nolte, S., Segev, M. & Szameit, A. Photonic Floquet topological insulators. *Nature* **496**, 196 (2013).

12. Khanikaev, A. B., Mousavi, S. H., Tse, W.-K., Kargarian, M., MacDonald, A. H. & Shvets, G. Photonic topological insulators. *Nat. Mater*. **12**, 233 (2013).

13. Wu, L.-H. & Hu, X. Scheme for Achieving a topological photonic crystal by using dielectric material. *Phys. Rev. Lett*. **114**, 223901 (2015).

14. Ozawa, T., Price, H. M., Amo, A., Goldman, N., Hafezi, M., Lu, L., Rechtsman, M. C., Schuster, D., Simon, J., Zilberberg, O. & Carusotto, I. Topological photonics. *Rev. Mod. Phys*. **91**, 015006 (2019).

15. Yang, Y., Xu, Y. F., Xu, T., Wang, H. X., Jiang, J. H., Hu, X. & Hang, Z. H. Visualization of a unidirectional electromagnetic waveguide using topological photonic crystals made of dielectric materials. *Phys. Rev. Lett.* **120**, 217401 (2018).

16. Barik, S., Karasahin, A., Flower, C., Cai, T., Miyake, H., DeGottardi, W., Hafezi, M. & Waks, E. A topological quantum optics interface. *Science* **359**, 666 (2018).

17. Parappurath, N., Alpeggiani, F., Kuipers, L. & Verhagen, E. Direct observation of topological edge states in silicon photonic crystals: Spin, dispersion, and chiral routing. *Sci. Adv*. **6**, eaaw4137 (2020).

18. Shao, Z. K., Chen, H. Z., Wang, S., Mao, X. R., Yang, Z. Q., Wang, S. L., Wang, X.-X., Hu, X. & Ma, R. M. A high-performance topological bulk laser based on band-inversion-induced reflection. *Nat. Nanotechnol.* **15**, 67 (2020).

19. Kagami, H., Amemiya, T., Okada, S., Nishiyama, N. & Hu, X. Highly efficient vertical coupling to a topological waveguide with defect structure. *Opt. Express* **29**, 32755 (2021).

20. Liu, W., Ji, Z., Wang, Y., Modi, G., Hwang, M., Zheng, B., Sorger, V. J., Pan, A. & Agarwal, R. Generation of helical topological exciton-polaritons. *Science* **370**, 600 (2020).

21. Yang, Z., Lustig, E., Harari, G., Plotnik. Y., Lumer, Y., Bandres, M. A. & Segev, M. Mode-Locked Topological Insulator Laser Utilizing Synthetic Dimensions. *Phys. Rev. X* **10**, 011059 (2020).

22. Guddala, S., Komissarenko, F., Kiriushechkina, S., Vakulenko, A., Li, M., Menon, V. M., Alù, A. & Khanikaev, A. B. Topological phonon-polariton funneling in midinfrared. *Science* **374**, 225 (2021).

23. Peano, V., Sapper, F. & Marquardt, F. Rapid exploration of topological band structures using deep learning. *Phys. Rev. X* **11**, 021052 (2021).

24. Dikopoltsev, A., Harder, T. H., Lustig, E., Egorov, O. A., Beierlein, J., Wolf, A., Lumer, Y., Emmerling, M., Schneider, C., Höfling, S., Segev, M. & Klembt, S. Topological insulator vertical-cavity laser array. *Science* **373**, 1514-1517 (2021).

25. Sun, X.-C., Wang, X.-X., Amemiya, T. & Hu, X. Comment on "Spin-momentum-locked edge mode for topological vortex lasing". *Phys. Rev. Lett*. **127**, 209401 (2021).

26. Sun, X.-C. & Hu, X. Topological ring-cavity laser formed by honeycomb photonic crystals, *Phys. Rev. B* **103**, 245305 (2021).

27. Cheng, X., Jouvaud, C., Ni, X., Mousavi, S. H., Genack, A. Z. & Khanikaev, A. B. Robust reconfigurable electromagnetic pathways within a photonic topological insulator. *Nat. Mater.* **15**, 542 (2016).

28. Qiao, Z., Jung, J., Lin, C., Ren, Y., MacDonald, A. H. & Niu, Q. Current Partition at Topological Channel Intersections. *Phys. Rev. Lett.* **112**, 206601 (2014).

29. Yan, M., Lu, J., Li, F., Deng, W., Huang, X., Ma, J. & Liu, Z. On-chip valley topological materials for elastic wave manipulation. *Nat. Mater.* **17**, 993 (2018).

30. Wu, X., Meng, Y., Tian, J., Huang, Y., Xiang, H., Han, D. & Wen, W. Direct observation of valley-polarized topological edge states in designer surface plasmon crystals. *Nat. Commun.* **8**, 1304 (2017).

31. Zhang, L., Yang, Y., He, M., Wang, H., Yang, Z., Li, E., Gao, F., Zhang, B., Singh, R., Jiang, J. & Chen, H. Valley kink states and topological channel intersections in substrate-integrated photonic circuitry. *Laser Photon. Rev.* **13**, 1900159 (2019).

32. Kang, Y., Ni, X., Cheng, X., Khanikaev, A. B. & Genack, A. Z. Pseudo-spin-valley coupled edge states in a photonic topological insulator. *Nat. Commun.* **9**, 3029 (2018).

33. Yang, Y., Qian, X., Shi, Li., Shen, X. & Hang, Z. Unidirectional transport in amorphous topological photonic crystals. *Sci. China Phys. Mech. Astron.* **66**, 274212 (2023).

34. Kariyado, T. & Hu, X. Topological states characterized by mirror winding numbers in graphene with bond modulation. *Sci. Rep.* **7**, 16515 (2017).

35. Barik, S., Miyake, H., DeGottardi, W., Waks, E. & Hafezi, M. Two-dimensionally confined topological edge states in photonic crystals. *New J. Phys.* **18**, 113013 (2016).

36. Wang, X.-X., Guo, Z., Song, J., Jiang, H., Chen, H. & Hu, X. Unique Huygens-Fresnel electromagnetic transportation of chiral Dirac wavelet in topological photonic crystal. *Nat. Commun.* **14**, 3040 (2023).

37. Jackiw, R. & Rebbi, C. Solitons with fermion number ½. *Phys. Rev. D* **13**, 3398 (1976).

38. Jackson, J. D. *Classical Electrodynamics* (Wiley, 1998).

39. Yablonovitch, E. Inhibited spontaneous emission in solid-state physics and electronics. *Phys. Rev. Lett.* **58**, 2059 (1987).

40. Sakoda, K. *Optical Properties of Photonic Crystals* (Springer, 2005).

41. Joannopoulos, J. D., Johnson, S. G., Winn, J. N. & Meade, R. D. *Photonic Crystals: Molding the Flow of Light* (Princeton University Press, 2008).

42. Huang, X., Lai, Y., Hang, Z. H., Zheng, H. & Chan, C. T. Dirac cones induced by accidental degeneracy in photonic crystals and zero-refractive-index materials. *Nat. Mater.* **10**, 582 (2011).

43. Haldane, F. D. M. Model for a quantum Hall effect without Landau levels: Condensed-matter realization of the "parity anomaly". *Phys. Rev. Lett.* **61**, 2015 (1988).

44. Hasan, M. Z. & Kane, C. L. Colloquium: Topological insulators. *Rev. Mod. Phys.* **82**, 3045 (2010).

45. Qi, X.-L. & Zhang, S.-C. Topological insulators and superconductors. *Rev. Mod. Phys.* **83**, 1057 (2011).

46. Weng, H., Yu, R., Hu, X., Dai, X. & Fang, Z. Quantum anomalous Hall effect and related topological electronic states. *Adv. Phys.* **64**, 227 (2015).

47. Gorlach, M. A., Ni, X., Smirnova, D. A., Korobkin, D., Zhirihin, D., Slobozhanyuk, A. P., Belov, P. A., Alù, A. & Khanikaev, A. B. Far-field probing of leaky topological states in all-dielectric metasurfaces. *Nat. Commun.* **9**, 909 (2018).

48. Palmer, S. J. & Giannini, V. Berry bands and pseudo-spin of topological photonic phases. *Phys. Rev. Res.* **3**, L022013 (2021).

49. Kane, C. L. & Mele, E. J. Quantum spin Hall effect in graphene. *Phys. Rev. Lett.* **95**, 226801 (2005).

50. Bernevig, B. A., Hughes, T. L. & Zhang, S.-C. Quantum spin Hall effect and topological phase transition in HgTe quantum wells. *Science* **314**, 1757 (2006).

51. COMSOL Multiphysics®v. 6.0. www.comsol.com. COMSOL AB, Stockholm, Sweden.

52. Siroki, G., Huidobro, P. A. & Giannini, V. Topological photonics: From crystals to particles. *Phys. Rev. B* **96**, 041408 (2017).

53. Berry, M. V. Quantal phase factors accompanying adiabatic changes. *Proc. R. Soc. Lond. A* **392**, 45 (1984).

54. Xiao, D., Chang, M.-C. & Niu, Q. Berry phase effects on electronic properties. *Rev. Mod. Phys.* **82**, 1959 (2010).

55. Chen, X., Lin, J. & Wang, K. A review of silicon-based integrated optical switches. *Laser Photonics Rev*. **17**, 2200571 (2023).

56. Kagami, H., Amemiya, T., Okada, S., Nishiyama, N. & Hu, X. Topological converter for high-efficiency coupling between Si wire waveguide and topological waveguide. *Opt. Express* **28**, 33619 (2020).

57. He, C., Ni, X., Ge, H., Sun, X.-C., Chen, Y.-B., Lu, M.-H., Liu, X.-P. & Chen, Y.-F. Acoustic topological insulator and robust one-way sound transport. *Nat. Phys.* **12**, 1124 (2016).

58. Cha, J., Kim, K. W. & Daraio, C. Experimental realization of on-chip topological nanoelectromechanical metamaterials. *Nature* **564**, 229 (2018).

## Acknowledgments

This work is supported by CREST, JST (Core Research for Evolutionary Science and Technology, Japan Science and Technology Agency) (Grant Number JPMJCR18T4) and partially by JPJSBP-120219942 (X.H.). M. Kawasaki is grateful for stimulating discussions.

## Author contributions

X.H. conceived and supervised the project. X.X.W. and X.H. performed the analytical calculations and wrote the manuscript. X.X.W. performed the computer simulations. T.A. joined discussions. All authors fully contribute to the research.

## Competing interests

The authors declare no competing financial and non-financial interests.

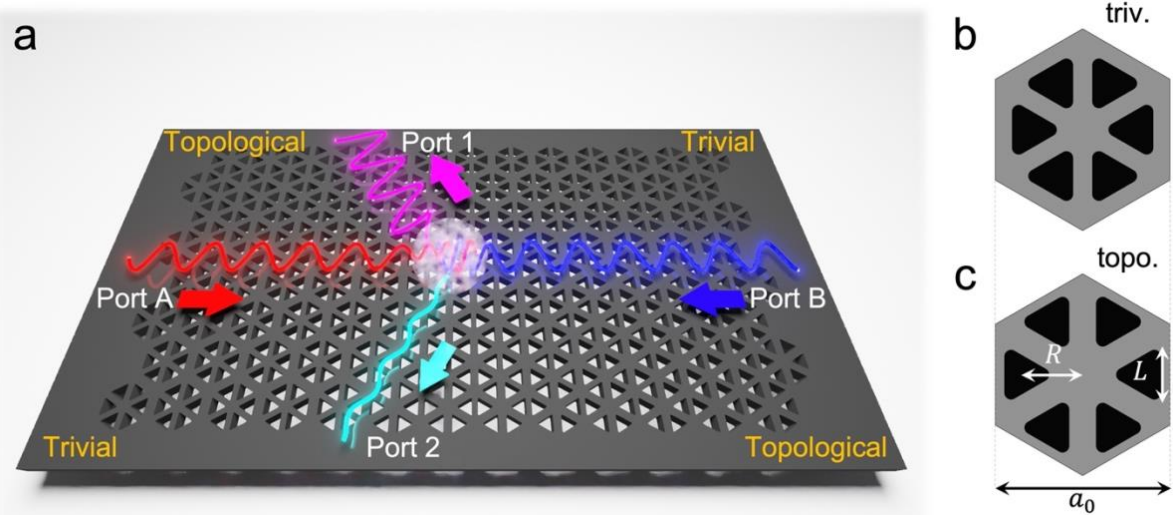

**Fig. 1 Schematic of the topological interferometry switch. a** Photonic crystal (PhC) based on a semiconductor membrane with triangular airholes constituted of two topological domains (upper left and lower right) and two trivial domains (upper right and lower left). The interfaces of the four domains are used as input (Port A and B) and output (Port 1 and 2) ports. Two input waves with a phase difference interfere right on the crossing point of the four topological interfacial waveguides and modulate the ratio of output powers at Port 1 and Port 2. **b** and **c** Structures of unit cells for the trivial and topological PhCs. The lattice constant $a_0$ and the side length of the triangular airholes $L$ are the same in both structures. $R$ is the distance measured from the center of unit cell to the centers of six airholes. One has $R < a_0/3$ in a unit cell for the trivial PhC and $R > a_0/3$ in a unit cell for the topological PhC.

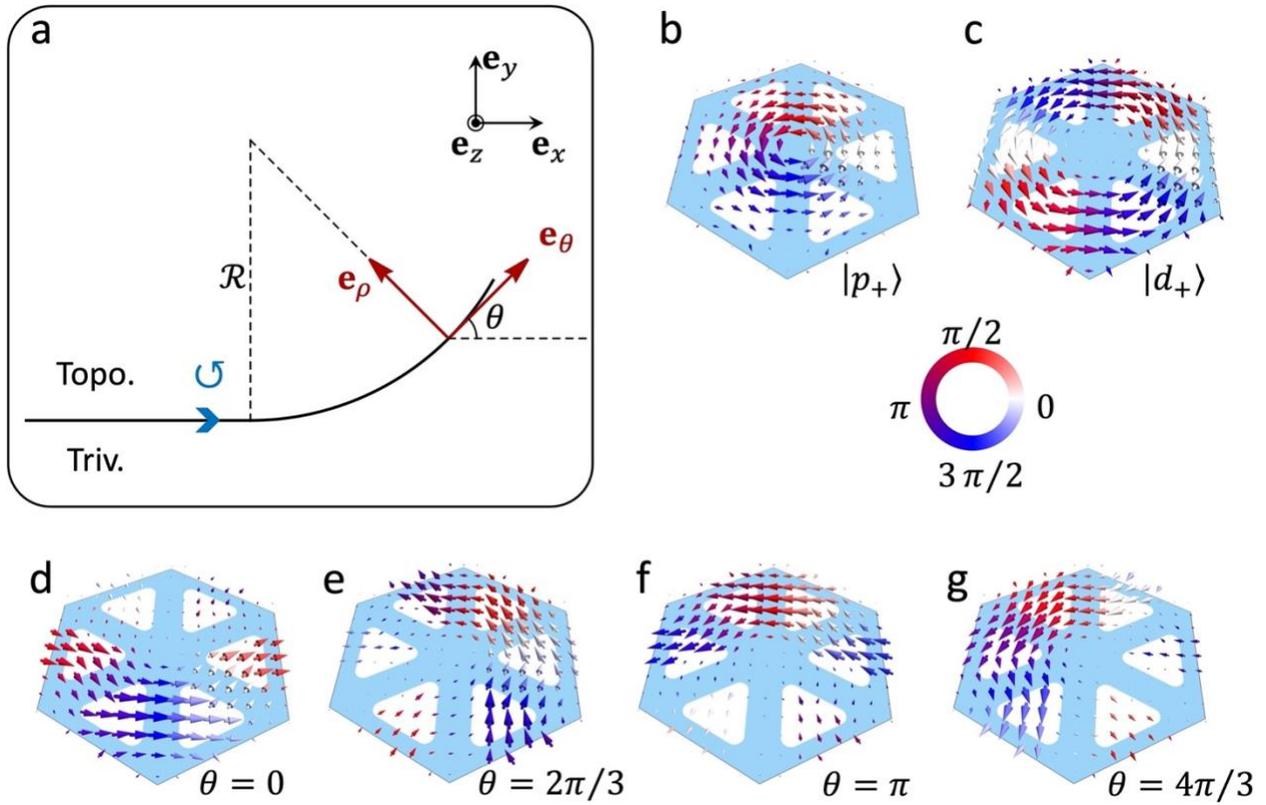

**Fig. 2 Transportation of pseudospin-up interfacial modes**. **a** Schematic of the polar-like coordinate for a topological interface bent with a fixed radius $\mathcal{R}$. **b** and **c** Pseudospin-up eigenmodes $|p_+\rangle$ and $|d_+\rangle$ at $\Gamma$ point as the eigen wavefunctions in a bulk system, where the arrows represent the local Poynting vectors and the color of which stand for the phase of wavefunction. **d-g** Pseudospin-up Jackiw-Rebbi modes $|\theta,+\rangle$ with $\theta = 0$, $\theta = 2\pi/3$, $\theta = \pi$ and $\theta = 4\pi/3$, respectively, which are obtained by superposing the wavefunctions of $|p_+\rangle$ and $|d_+\rangle$ in (**b**) and (**c**) with relative phase factors associated with the directions of topological interfacial waveguides.

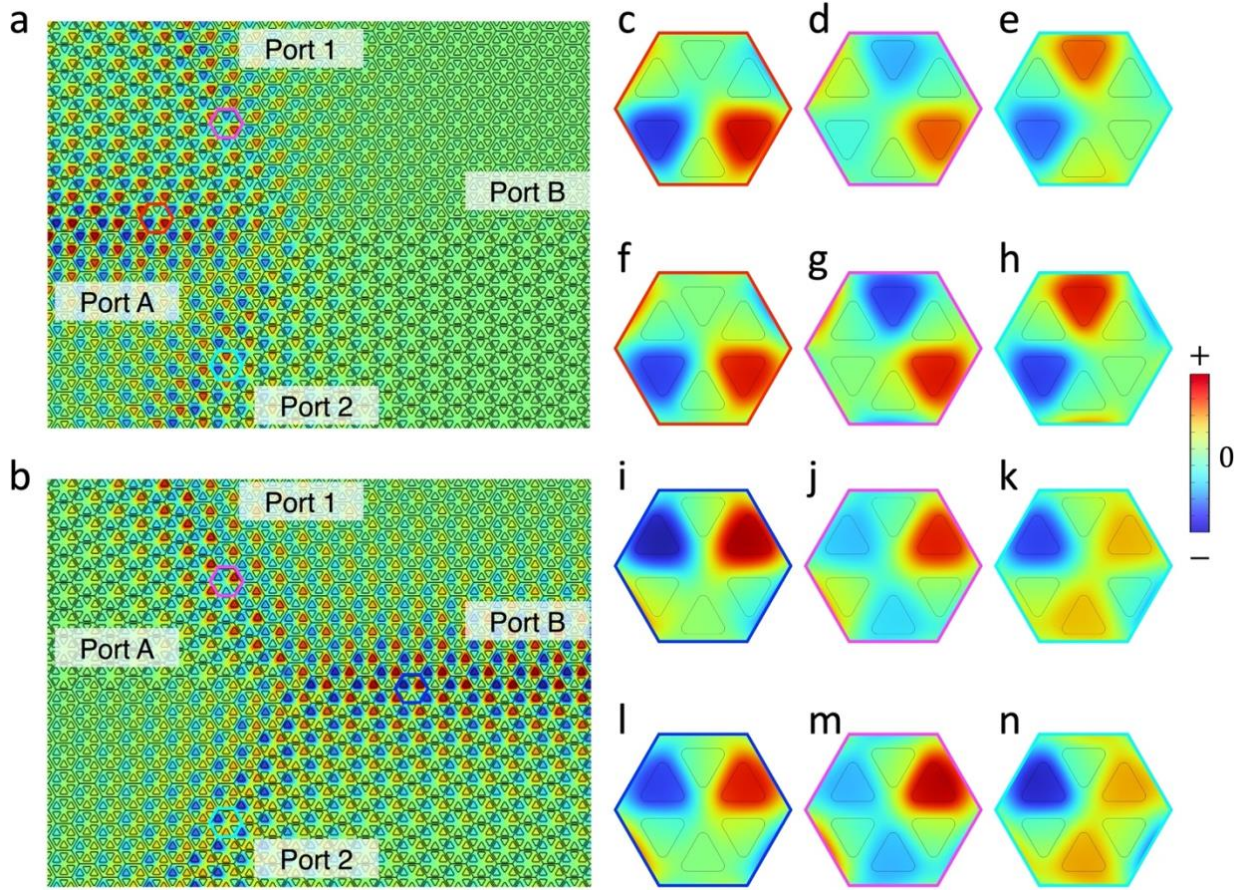

**Fig. 3 Splitting of topological interfacial mode obtained by simulations when one of the two input ports is activated. a** Real part of the simulated wavefunction $H_z$ (out-of-plane magnetic field), where only Port A is activated. **b** Same as (**a**) except that only Port B is activated. **c-e** Zoom-in view of wavefunction inside a typical unit cell encompassed by color hexagon at Port A, Port 1 and Port 2, respectively, in (**a**). The corresponding unit cells in (**a**) and (**c-e**) are marked by hexagonal frames with the same colors. **f-h** Wavefunctions constructed by superposing $|p_+\rangle$ and $|d_+\rangle$ using the expressions of $|A\rangle$, $|1\rangle_A$ and $|2\rangle_A$ obtained theoretically according to Eqs. (10) and (13), which match well with those in (**c-e**). **i-n** Same as (**c-h**) except that only Port B is activated. Color looks different for the real parts of wavefunctions in (**a**) and (**b**) due to different phase accumulations while the whole intensity of EM wave is the same.

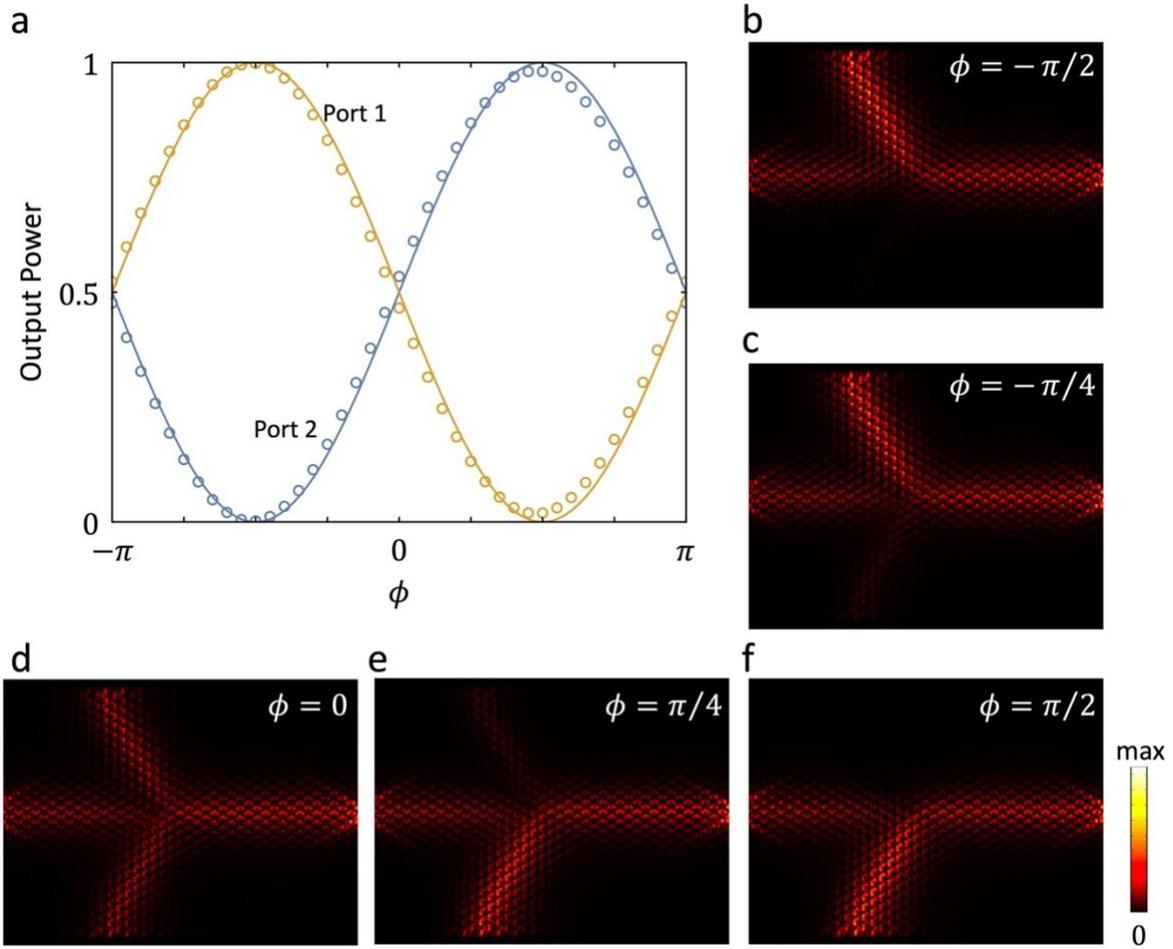

**Fig. 4 Interferometry switching phenomena when two input ports are activated simultaneously with finite phase difference. a** Proportion of the total output power at Port 1 (yellow) and Port 2 (blue) as function of the phase difference $\phi$ between input Port B and Port A. Circle marks are for simulated results, whereas the solid lines are for the theoretical results given by Eq. (16). **b-f** Simulated distribution of the field strength $|H_z|^2$ for $\phi = -\pi/2$, $\phi = -\pi/4$, $\phi = 0$, $\phi = \pi/4$ and $\phi = \pi/2$, respectively.